\documentclass[aps,prb,preprint,]{revtex4-1}
\usepackage{amssymb}
\usepackage{graphicx}
\usepackage{amsmath}
\usepackage{amsfonts}
\usepackage{color}

\begin{document}

\title{Non-Hermitian semi-Dirac semi-metals}

\author{Ayan Banerjee}
\affiliation{Solid State and Structural Chemistry Unit, Indian Institute of Science, Bangalore 560012, India}
\author{Awadhesh Narayan}
\email{awadhesh@iisc.ac.in}
\affiliation{Solid State and Structural Chemistry Unit, Indian Institute of Science, Bangalore 560012, India}

\date{\today}

\begin{abstract}
Recently, many novel and exotic phases have been proposed by considering the role of topology in non-Hermitian systems, and their emergent properties are of wide current interest. In this work we propose the non-Hermitian generalization of semi-Dirac semimetals, which feature a linear dispersion along one momentum direction and a quadratic one along the other. We study the topological phase transitions in such two-dimensional semi-Dirac semimetals in the presence of a particle gain-and-loss term. We show that such a non-Hermitian term creates exceptional points originating out of each semi-Dirac point. We map out the topological phase diagram of our model, using winding number and vorticity as topological invariants of the system. By means of numerical and analytical calculations, we examine the nature of edge states for different types of semi-Dirac models and establish bulk-boundary correspondence and absence of the non-Hermitian skin effect, in one class. On the other hand, for other classes of semi-Dirac models with asymmetric hopping, we restore the non-Hermitian skin effect, an anomalous feature usually present in non-Hermitian topological systems.
\end{abstract}

\maketitle
\section{Introduction}

Over the last three decades, the studies of topological phases~\cite{PhysRevLett.45.494,PhysRevLett.49.405,haldane1988model,PhysRevLett.95.226801,Bernevig1757}, including topological insulators~\cite{RevModPhys.88.035005,RevModPhys.83.1057,RevModPhys.82.3045}, Chern insulators ~\cite{haldane1988model,qi2006topological,neupert2011fractional}, topological superconductors~\cite{PhysRevLett.100.096407,RevModPhys.83.1057}, and topological semimetals~\cite{armitage2018weyl,kruthoff2017topological,doi:10.1146/annurev-conmatphys-031016-025458,bradlyn2016beyond}, have been becoming a growing topic of interest in condensed matter physics. Among various topological systems the search for gapless yet topological phases, such as Dirac semimetals and Weyl semimetals, and more recently nodal line and multifold semimetals, is sprouting into fascinating new directions~\cite{armitage2018weyl}.

In quantum mechanics every physical operator is represented by a Hermitian operator and hence one obtains real eigen spectra, and at the same time conservation of probabilities~\cite{dirac1981principles}. Yet, in recent years, Bender and  co-workers have demonstrated  that  parity-time ($PT$) symmetric non-Hermitian Hamiltonians can show real spectra~\cite{bender1998real}. Considerable efforts have been devoted to study non-Hermitian phases in a variety of platforms, including open quantum systems~\cite{Rotter_2015,rotter2009non}, incorporating electron-phonon interactions~\cite{kozii2017nonhermitian,Yoshida_2018} and in quantum optics~\cite{ruter2010observation,refId0,PhysRevLett.101.080402,PhysRevLett.103.093902}. 

The research in this field has rapidly accelerated by considering the interplay between non-Hermiticity and topology~\cite{alvarez2018topological,borgnia2019non,lee2019anatomy,ghatak2019new,li2019geometric,torres2019perspective,yoshida2019non,bergholtz2019exceptional,yoshida2019mirror}. Several non-Hermitian non-trivial topological phases have been proposed, including one-dimensional Su-Schrieffer-Heeger chains~\cite{jin2017schrieffer,PhysRevA.98.013628,PhysRevA.98.023808}, knot semimetals~\cite{carlstrom2018exceptional,carlstrom2019knotted}, nodal line semimetals, nodal ring semimetals~\cite{wang2019non,yoshida2019symmetry}, Hopf link semimetals~\cite{PhysRevB.99.041406,PhysRevLett.118.045701,PhysRevB.99.041116,PhysRevB.99.081102,lee2018tidal}, and Weyl semimetals~\cite{xu2017weyl}, to name just a few. One of the features of non-Hermitian topological systems is the existence of exceptional points (EPs). EPs are singularities where both eigenvalues and eigenvectors coalesce and result in the Hamiltonian becoming "defective"~\cite{heiss2012physics}. Encircling such an EP yields a quantized topological invariant. Along with theoretical developments, many ingenious experiments have been performed recently in ongoing efforts to engineer non-Hermitian systems. Some of the notable ones among them are experiments on microwave cavities~\cite{poli2015selective,midya2018non}, lossy waveguides~\cite{zeuner2015observation,weimann2017topologically,chen2014experimental}, topological lasers~\cite{harari2018topological,feng2014single,bandres2018topological,zhao2018topological,parto2018edge}, topolectrical circuits~\cite{helbig2019observation,li2019topology} and in quantum optics setups~\cite{ozawa2019topological,rechtsman2013photonic,goldman2016topological}.

An intriguing and unusual type of semimetal is one which has a linear dispersion along one momentum direction and a quadratic dispersion along another. Systems with such peculiar dispersions have been termed semi-Dirac semimetals. A number of candidate hosts have been proposed. These include honeycomb and square lattices under a magnetic field~\cite{dietl2008new,delplace2010semi}, transition metal oxide heterostructures~\cite{pardo2009half}, as well as photonic crystals~\cite{wu2014semi}. Several interesting properties of semi-Dirac semimetals, which are distinct from other semimetals, have also been subsequently revealed~\cite{banerjee2009tight,banerjee2012phenomenology,narayan2015,saha2016photoinduced}.

In this paper, we introduce the generalization of semi-Dirac semimetals to the non-Hermitian case. We present both continuum and lattice models of non-Hermitian semi-Dirac semimetals in the presence of gain and loss terms, and show that a new topological phase arises on introducing such a non-Hermitian term. We use analytical as well as numerical calculations to illustrate the topological features of the non-Hermitian semi-Dirac system. Employing two different topological invariants, namely vorticity and winding number, we map out the phase diagram of the system. Using computations under open boundary conditions, we examine the bulk-boundary correspondence and reveal the absence of non-Hermitian skin effect in one class of semi-Dirac models. We also show that in other classes of models with asymmetric hopping, the non-Hermitian skin effect is present. Our hope is that these findings will motivate future theoretical and experimental investigations of non-Hermitian semi-Dirac systems.

\section{Results}

\begin{figure*}
\includegraphics[scale=0.30]{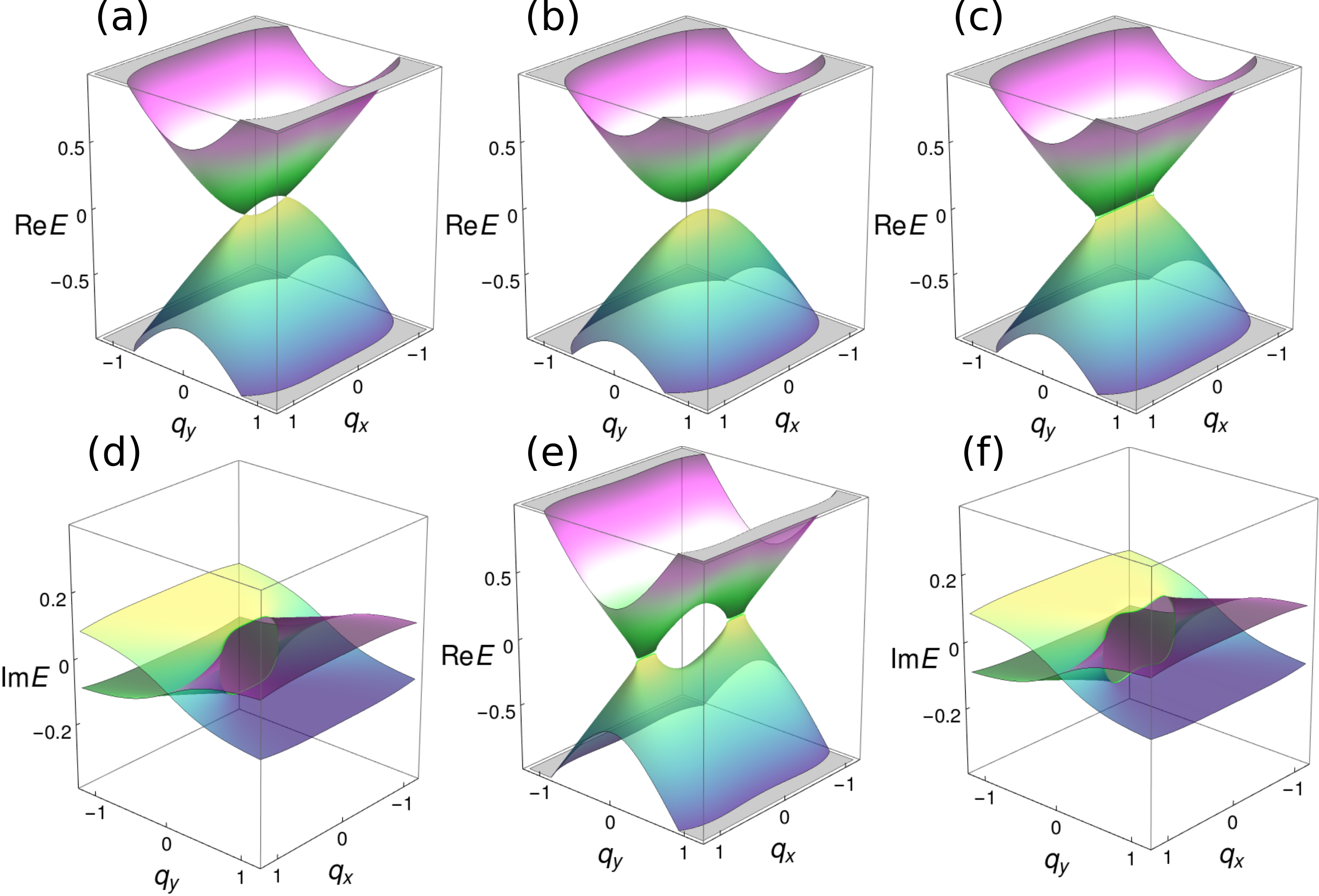}
  \caption{\textbf{Band diagrams of the non-Hermitian semi-Dirac model.} (a) Two gap-less semi-Dirac points for $\delta_0=0.05$ in the absence of non-Hermitian term ($\gamma=0.0$). (b) Energy spectrum is gapped for negative $\delta_0$. Here we choose $\delta_0=-0.05,\gamma=0.0$. (c) Real part of the energy with $\delta_0=0$. (d) Imaginary part of the energy with $\delta_0=0$. (e) Real part of the energy with nonzero $\delta_0=0.25$. (f) Imaginary part of the energy with nonzero $\delta_0=0.25$. For (c)-(e) we include the non-Hermitian term $i\gamma\sigma_z$ ($\gamma=0.1$). We get two EPs in (c) and four EPs in (e) along $q_y=0$ line. The imaginary eigenvalues are interchanged along $q_y=0$ line in (d) and (f). For the other parameters, we choose the following values: $m=v_f=1.0$, keeping them unchanged unless otherwise specified.} \label{band_diagram}
\end{figure*}

\begin{figure*}
\includegraphics[scale=0.28]{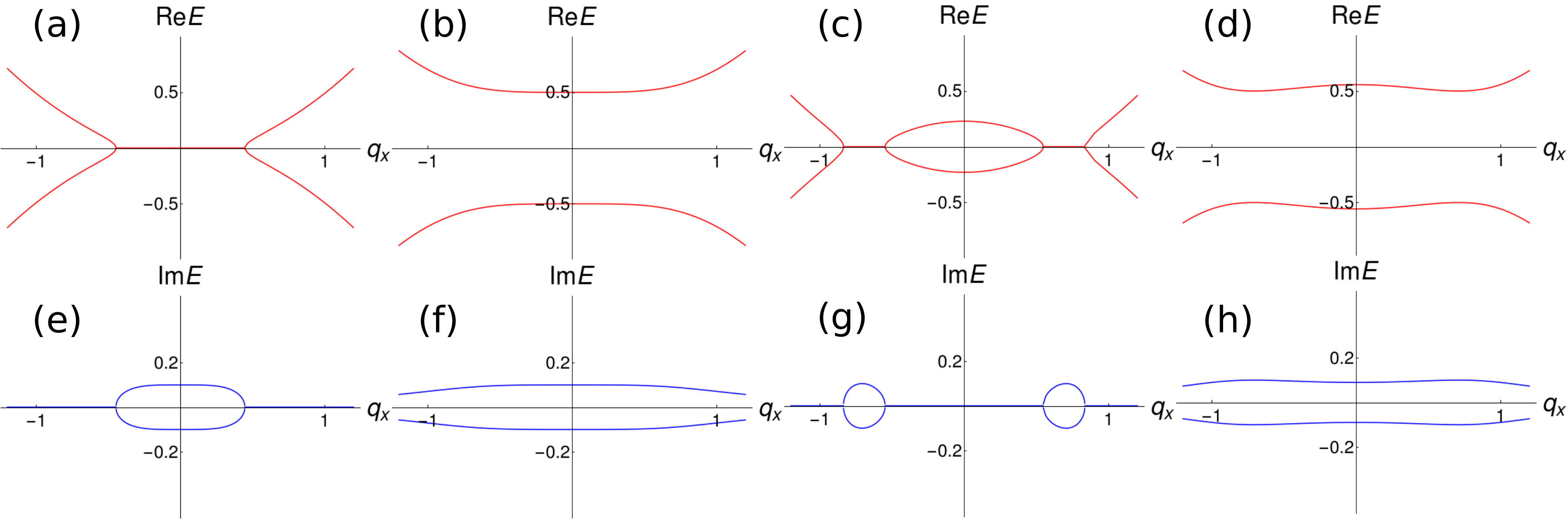}
  \caption{\textbf{Band diagrams of the non-Hermitian semi-Dirac model showing PT symmetric real eigenvalues with $q_y$ zero and non-zero.} Two exceptional points arising from gap-less semi-Dirac points for $\delta_0=0$ in the presence of a finite non-Hermitian term ($\gamma=0.1$) are shown in (a) with real part and (e) imaginary part of eigenvalues with $q_y=0$. (b) Real and (f) imaginary part of energy spectrum is shown for $\delta_0=0$ with $q_y$ nonzero. (c) Real part of the energy with $\delta_0=0.25$. (g) Imaginary part of the energy with $\delta_0=0.25$ and $q_y=0$. (d) Real and (h) Imaginary part of the energy with nonzero $\delta_0=0.25$ for $q_y$ nonzero. We get two EPs in (a) and (e) and four EPs in (c) and (g) along the $q_y=0$ line. The spectra are gapped for finite $q_y$. We see PT symmetric real eigenvalues in the presence of non-Hermitian term along $q_y=0$ line. We set $q_y=0.5$ as nonzero $q_y$ value.} \label{PT_symmetric band_diagram}
\end{figure*}

This section is organized in the following manner. We will first present the low-energy model for semi-Dirac semimetals and the extension to the non-Hermitian case. We will then present the calculation of the winding number and vorticity, which enables mapping out the topological phase diagram. Finally we will examine the bulk-boundary correspondence in this system.

\subsection{Model}

The model Hamiltonian describing low-energy electronic bands of a two-dimensional semi-Dirac semi-metal(More precisely this should be termed a semi-Weyl dispersion since we are using a two-band model.) is~\cite{delplace2010semi,saha2016photoinduced,banerjee2009tight}

\begin{equation}  \label{Hamiltonian}
 H_0(\textbf q)=\textbf d(\textbf q)\cdot\boldsymbol {\sigma},
\end{equation}

\noindent where $\boldsymbol{\sigma}= (\sigma_x, \sigma_y, \sigma_z)$ are the Pauli matrices in the pseudospin space and $\textbf d(\textbf q) = (\frac{q_x^2}{2m}-\delta_0, 0,v_fq_y)$. Here $\textbf q=(q_x,q_y)$ is the crystal momentum, $m$ is the quasi particle mass along the quadratically dispersing direction, $v_f$ is the Dirac velocity, and $\delta_0$ is the gap parameter. The band gap is tuned by the gap parameter $\delta_0$. For $\delta_0 >0$ we have two gapless Dirac points, as shown in Fig.~\ref{band_diagram}(a). We obtain a fully gapped trivial insulator for $\delta_0<0$ [see Fig.~\ref{band_diagram}(b)]. In the intermediate case of $\delta_0=0$, the spectrum is gapless with a semi-Dirac dispersion. For finite $\delta_0$, the overall dispersion differs from that of isolated Dirac points, due to the curvature effect arising from the quadratic momentum term and leading to the saddle point. However, $\delta_0$ controls the Fermi surface topology via Lifshitz transitions. The $PT$ symmetry operator~\cite{zhang2016quantum} for our Hermitian semi-Dirac semimetal system can be represented as the complex conjugation $K$ in spinless orbital basis which guarantees the real eigenspectra, i.e., $H(\mathbf{k}) = H(\mathbf{k})^\ast$. Now, in addition, we introduce a non-Hermitian term to the original Hamiltonian (Eq.~\ref{Hamiltonian}) such that the Hamiltonian becomes
 
\begin{equation} \label{Non-Hermitian_Hamiltonian}
H(\textbf q)=\left(\frac{q_x^2}{2m}-\delta_0\right)\sigma_x+v_fq_y\sigma_z+i\gamma\sigma_z.
\end{equation}

The non-Hermitian perturbing term can be thought of as a gain and loss between the two orbitals, with $\gamma$ as the gain and loss coefficient~\cite{ruter2010observation}. The non-Hermitian perturbation explicitly breaks the $PT$ symmetry. Nevertheless, in some parameter regions, we do get real eigenvalues as a consequence of $PT$-like symmetry~\cite{lee2016anomalous,alvarez2018non,wang2019non}. We have shown different combinations with $q_y$ and $\delta_0$ zero and non-zero. It is clear from Fig.~\ref{PT_symmetric band_diagram} that the model shows PT symmetric real eigenvalues with $q_y=0$ even in the presence of $\gamma$ term. We get two and four EPs  with $\delta_0$ zero and non-zero respectively. We get gapped spectra for non-zero $q_y$. We then obtain the energy eigenvalues as

\begin {equation} \label{energy_eigenvalues}
E=\pm\sqrt{\left(\frac{q_x^2}{2m}-\delta_0\right)^2+v_f^2q_y^2-\gamma^2+2iv_fq_y\gamma},
\end{equation}
 
\noindent which is in general complex. As a consequence of non-Hermitian band degeneracy~\cite{Heiss_2012,rotter2009non,PhysRevLett.118.093002}, we expect the appearance of EPs. At these EPs not only the eigenvalues but also the eigenvectors coalesce rendering the Hamiltonian non-diagonizable~\cite{heiss2012physics}. Indeed, for our model in the parameter range $0<\gamma<\delta_0$, we discover a nodal line along $q_y=0$, where the four EPs exist. For the values of $\delta_0>0$, the two gap-less semi-Dirac points (for $\gamma=0$) get converted into four EPs (for $\gamma>0$), as can be seen in Fig.~\ref{band_diagram}(e). Examining the imaginary part of the spectrum, we find that the imaginary part of the energies are interchanged across the EP, as presented in Fig.~\ref{band_diagram}(d) and (f). Interestingly, a competition between the gap parameter $\delta_0$ and the non-Hermitian term $\gamma$ leads to annihilation or creation of EPs. Out of the four EPs, two of them annihilate each other at the critical value of $\gamma=\delta_0$. For the special case of $\delta_0=0$ and a nonzero $\gamma$, the gapless semi-Dirac points turn into two EPs, as is presented in Fig.~\ref{band_diagram}(c). The corresponding imaginary part of the energy is shown in Fig.~\ref{band_diagram}(d).

The locations of the four symmetrically placed EPs, lying along the line $q_y=0$, are

\begin{equation} \label{location_EP}
q_{x}^{{EP}}=\pm \sqrt{2m(\delta_0\pm\gamma)}.
\end{equation}

Without the loss of generality, we now consider the case with $\delta_0>0$. As a consequence of the square root singularity in the complex energy spectrum~\cite{berry2010geometric}, we can expect topological phase transitions around the EPs. Next, we will calculate the winding number ~\cite{leykam2017edge,zhou2018dynamical}, which is closely related to the non-Hermitian generalization of the Berry phase~\cite{garrison1988complex,leykam2017edge}, in order to monitor and characterize topological phase transitions in our model. 
In this work, we have used two different topological invariants in order to study the bulk topology from different perspectives. EPs are spectral degenerate points in non-Hermitian eigenspectra, where not only eigenvalues but also eigenvectors coalesce with each other, resulting in the corresponding Hamiltonian becoming defective. Vorticity records how the constitutive bands get exchanged while encircling an EP owing to square root singularity in the dispersion~\cite{PhysRevLett.120.146402}. In the context of topological phase transitions, the calculation of vorticity does not give a complete picture as the loop enclosing two EPs with opposite vorticities and the loop enclosing no EPs give the same zero vorticity. Moreover, it depends only on energy eigenvalues. On the other hand, the construction of winding number is based on eigenstates. It originates from the chiral symmetry of the non-Hermitian Hamiltonian based on the non-Hermitian generalization of the Berry phase ~\cite{wang2019non,yin2018geometrical}. Both of these topological numbers are related to the edge states under open boundary conditions and are intimately connected to the notion of bulk-boundary correspondence (or its lack thereof)~\cite{esaki2011edge}.

\subsection{Winding Number}

\begin{figure}
\includegraphics[scale=0.4]{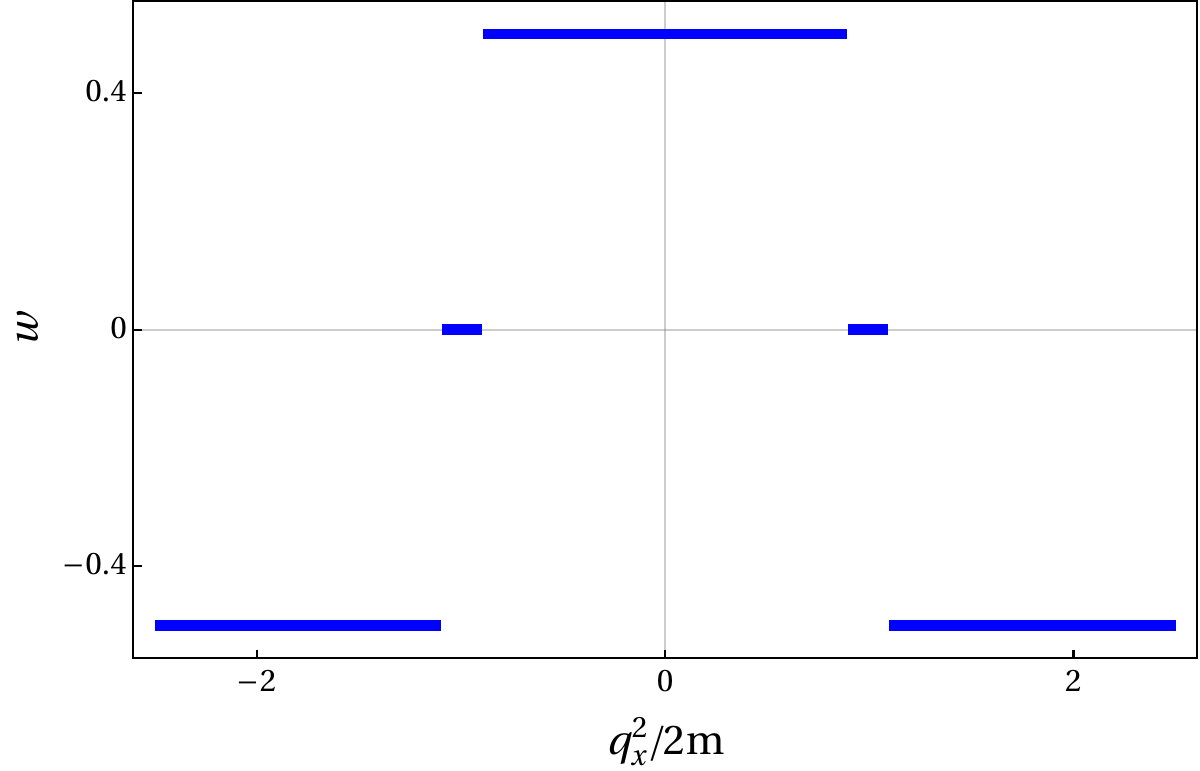}
  \caption{\textbf{Topological phase diagram from the winding number.} The winding number as a function of $q_x$ in the presence of the non-Hermitian term. We find three distinct topological phases with winding numbers $0$, $1/2$ and $-1/2$, respectively. Here we choose $\gamma=0.1$ and $\delta_0=0.5$.} \label{Topological_phase_Transition}
\end{figure}

In this sub-section, we will explicitly calculate the winding number for our model, which allows characterizing the topological phases. The winding number, $w$, is defined as~\cite{wang2019non,yin2018geometrical}

\begin{equation} \label{winding_number}
w= \frac{1}{2\pi}\int\limits_ {-\infty}^{\infty} \ dq_y\partial_{q_y} \ \phi.
\end{equation}

\noindent In the above expression, we define the winding number in the parameter space of the Hamiltonian along the $q_y$ direction. We express the Hamiltonian as

\begin{equation} \label{Hamiltonian_sgma-basis}
H=h_x\sigma_x+h_z\sigma_z,    
\end{equation}
       
\noindent with $h_x=(q_x^2/2m-\delta_0)$, $h_z=v_fq_y+i\gamma$, and $\phi$ defined as $\phi=\tan^{-1}(h_x/h_z)$. We obtain

\begin{equation} \label{Phi}
\phi=\tan^{-1}\left( \frac{q_x^2/2m -\delta_0}{v_fq_y+i\gamma}\right).
\end{equation}

As a complex angle, $\phi$ can be decomposed as $\phi=\phi_R+i\phi_I$, where $\phi_R$ and $\phi_I$ denote the real and imaginary parts of $\phi$, respectively. The values of $\phi$ for the limits $q_y\rightarrow \pm\infty$ are obtained as

\begin{equation} \label{phi_q_y_infinity-limit}
\begin{split}
\phi (q_y\rightarrow \pm \infty) & =\lim_{q_y\rightarrow \pm \infty} \tan^{-1}\left(\frac{q_x^2/2m -\delta0}{v_fq_y+i\gamma}\right)\\
&=\pm 0,
\end{split}
\end{equation}
     
\noindent which are purely real. Now using the relations

\begin{equation} \label{exponential_phi}
e^{2i\phi}=\frac{\cos\phi+i\sin\phi}{\cos\phi-i\sin\phi}=\frac{1+i\tan\phi}{1-i \tan\phi}=\frac{h_z+ih_x}{h_z-ih_x},
\end{equation}

\noindent we find that the amplitude and the phase are related to $\phi_I$ and $\phi_ R$ as~\cite{wang2019non,yin2018geometrical} 

\begin{equation} \label{exponential_phi_imaginary}
e^{-2\phi_I}= {\left|\frac{h_z+ih_x}{h_z-ih_x}\right|},
\end{equation}

\begin{equation} \label{exponential_phi_Real}
e^{2i\phi_R}=\frac{\frac{h_z+ih_x}{h_z-ih_x}} {\left|\frac{h_z+ih_x}{h_z-ih_x}\right|}.
\end{equation}

\begin{figure}
\includegraphics[scale=0.4]{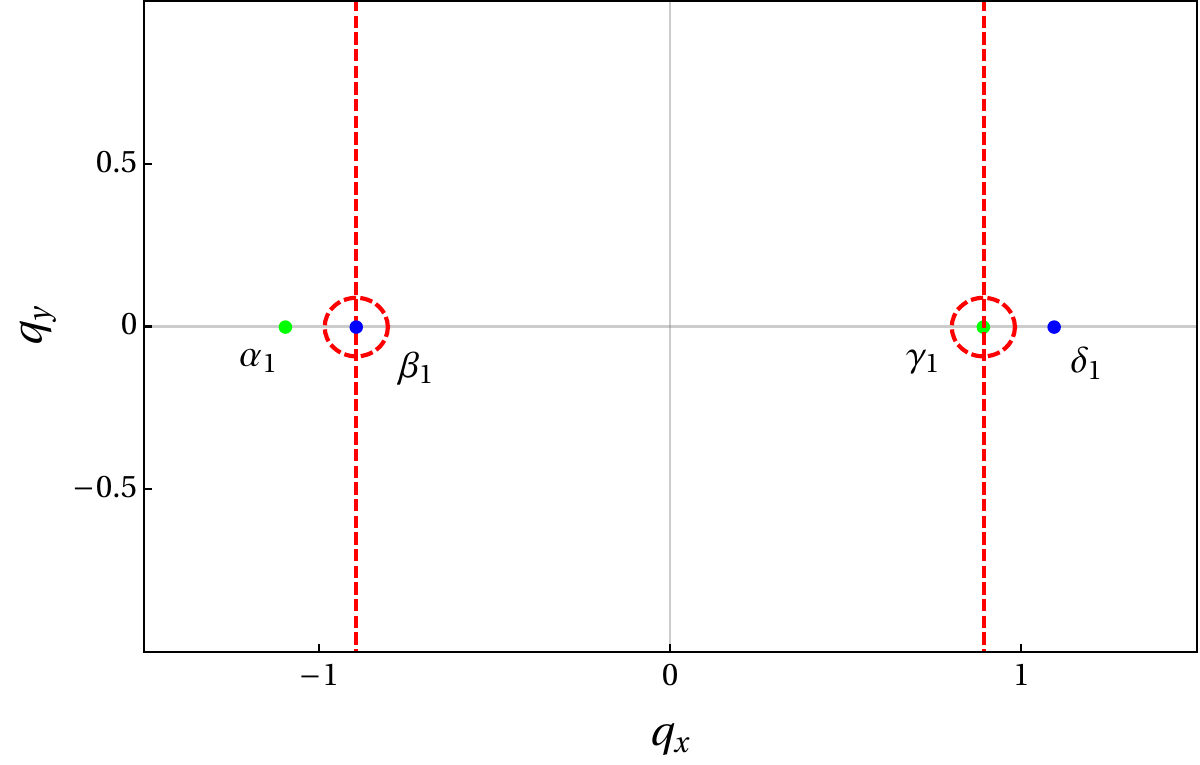}
  \caption{\textbf{Location of the exceptional points.} The four EPs are represented by blue and green dots. We designate them as $\alpha_1$, $\beta_1$, $\gamma_1$ and $\delta_1$. The green and blue dots correspond to the $1/2$ and $-1/2$ vorticities, respectively. The dotted line passing through EPs are representative of closed loops encircling each EP in the momentum space using periodic boundary conditions. These closed loops are treated as contours for calculating the winding number. Here we set $\gamma=0.1$ and $\delta_0=0.5$.}  \label{Location_of_EP}
\end{figure}

Now from Eq.~\ref{Phi}, we observe that $\phi_R$ is an odd function of $q_y$. Along $q_y=0$ line, it is discontinuous at each of the EPs and continuous between two successive EPs (see Fig.~\ref{Location_of_EP}). On the other hand, the real part of $\partial_{q_y}\phi$ is always continuous. In contrast, $\phi_{I}$ is an even and continuous function of $q_y$. So, $\partial_{q_y}\phi_I$ should be an odd function of $q_y$ and $\phi_I(q_y\rightarrow\infty)=\phi_I(q_y\rightarrow -\infty)$. The imaginary part of the integral in Eq.~\ref{winding_number} is obtained as

\begin{equation} \label{Imaginary_phi_integral}
\frac{1}{2\pi} \int\limits_ {-\infty}^{\infty} \ dq_y\partial_{q_y} \ \phi_I=\frac{\phi_I(q_y\rightarrow\infty)-\phi_I(q_y\rightarrow -\infty)}{2\pi}=0.
\end{equation}

Now, considering the relation 

\begin{equation} \label{Tangent_real_phi}
\tan(2\phi_R)=\frac{\mathrm{Im}\left(\frac{h_z+ih_x}{h_z-ih_x}\right)}{\mathrm{Re}\left(\frac{h_z+ih_x}{h_z-ih_x}\right)},
\end{equation}

\noindent we can express

\begin{equation} \label{Tangent_real_phi_interms_of_real_angles}
\tan(2\phi_R)=\tan(\phi_A+\phi_B),
\end{equation}

\noindent and

\begin{equation}\label{tangent_pi_A}
\tan\phi_A=\frac{\mathrm{Re}(h_x)+\mathrm{Im}(h_z)}{\mathrm{Re}(h_z)-\mathrm{Im}(h_x)}=\frac{(q_x^2/2m-\delta_0)+\gamma}{v_fq_y},
\end{equation}
\begin{equation} \label{tangent_pi_B}
\tan\phi_B=\frac{\mathrm{Re}(h_x)-\mathrm{Im}(h_z)}{\mathrm{Re}(h_z)+\mathrm{Im}(h_x)}=\frac{(q_x^2/2m-\delta_0)-\gamma}{v_fq_y}.
\end{equation}

In the above expressions, $\phi_A$ and $\phi_B$ are real angles. Simplifying these expressions, we obtain

\begin{equation} \label{phi_R_interms_of_phi_A_phi_B}
\phi_R=n\pi+\frac{1}{2}(\phi_A+\phi_B),
\end{equation}

\noindent where $n$ is an integer. We further arrive at the following equalities

\begin{equation} \label{phi_A_q-y_limit_0}
\phi_A(q_y\rightarrow0^\pm)=\pm\frac{\pi}{2} \mathrm{sgn}(\frac{q_x^2}{2m}-\delta_0+\gamma),
\end{equation}

\begin{equation} \label{phi_B_q-y_limit_0}
\phi_B(q_y\rightarrow0^\pm)=\pm\frac{\pi}{2} \mathrm{sgn}(\frac{q_x^2}{2m}-\delta_0-\gamma).
\end{equation}

So, we observe that both $\phi_A$ and $\phi_B$ have discontinuities at $q_y=0$. However, when $q_y\rightarrow\pm\infty$, we obtain $\phi_A(q_y\rightarrow\pm\infty)=\phi_B(q_y\rightarrow\pm\infty)=\pm0$. Now using Eqs.~\ref{winding_number},~\ref{phi_q_y_infinity-limit},~\ref{phi_A_q-y_limit_0},~\ref{phi_B_q-y_limit_0}, we can calculate the winding number as follows

\begin{equation} \label{Integration_winding_number}
\begin{split}
 w & = \frac{1}{2\pi}\int\limits_ {-\infty}^{\infty} \ dq_y\partial_{q_y}  \phi_R\\ 
 &  = \frac{1}{4\pi}\int\limits_ {-\infty}^{\infty} \ dq_y\partial_{q_y}(\phi_A+\phi_B)\\
 &  =\dfrac{1}{4\pi}\left((\phi_A|_{0^+} ^{+\infty}+\phi_A|^{0^-} _{-\infty} )+(\phi_B|_{0^+} ^{+\infty}+\phi_B|^{0^-} _{-\infty} )\right)\\
 & = -\frac{\mathrm{sgn}\left((\frac{q_x^2}{2m}-\delta_0)+\gamma\right)+\mathrm{sgn}\left((\frac{q_x^2}{2m}-\delta_0)- \gamma\right)}{4}. 
\end{split}
\end{equation}
  
Upon final simplification, we obtain the winding number
 
\begin{equation} \label{Final-winding-number}
\begin{split}
w & =-\frac{1}{2}, \quad  {{q_x^2/2m  > \delta_0+\gamma}}\\
& = 0, \quad  \delta_0-\gamma<\frac{q_x^2}{2m}<\delta_0+\gamma\\
& = \frac{1}{2},\quad  0<q_x^2/2m<\delta_0-\gamma.
\end{split}
\end{equation}

So, we find that for our semi-Dirac model there are three topologically distinct regions with winding numbers $w=\pm 1/2,0$. We present the topological phase diagram in Fig.~\ref{Topological_phase_Transition}. Interestingly, in this model the winding number acquires fractional values, as can also be realized in non-Hermitian Dirac~\cite{rui2019pt} and line nodal semimetals~\cite{wang2019non}. We can interpret the fractional value of $w$ as follows: As both the values of $\phi$ and its derivative, $\partial_y\phi$, are continuous along the $q_y$ line (from $q_y\rightarrow -\infty$ to $q_y\rightarrow \infty$), we can convert the line integral extended along $q_y$ to a loop using periodic boundary conditions, in a spirit similar to the Bloch Hamiltonian in lattice models (see Fig.~\ref{Location_of_EP}). We can then create loops encircling each of the EPs. These loops are topologically equivalent to the infinitely extended line along $q_y$. Whenever the loop encircles an EP, the winding number is found to be $\pm \frac{1}{2}$, as pointed out by Lee in Ref.~\onlinecite{lee2016anomalous}, indicating different topological phases for two consecutive EPs. We get three distinct regions with different winding number. A complementary point of view is obtained by considering the winding number of the parent Hermitian semi-Dirac model, which is zero~\cite{banerjee2012semi}. On introducing a non-Hermitian term in the Hamiltonian, each of the semi-Dirac point splits into two EPs, each with a winding number of $\pm 1/2$, thereby conserving the total winding number~\cite{lin2019symmetry}.

\begin{figure*}
\includegraphics[scale=0.5]{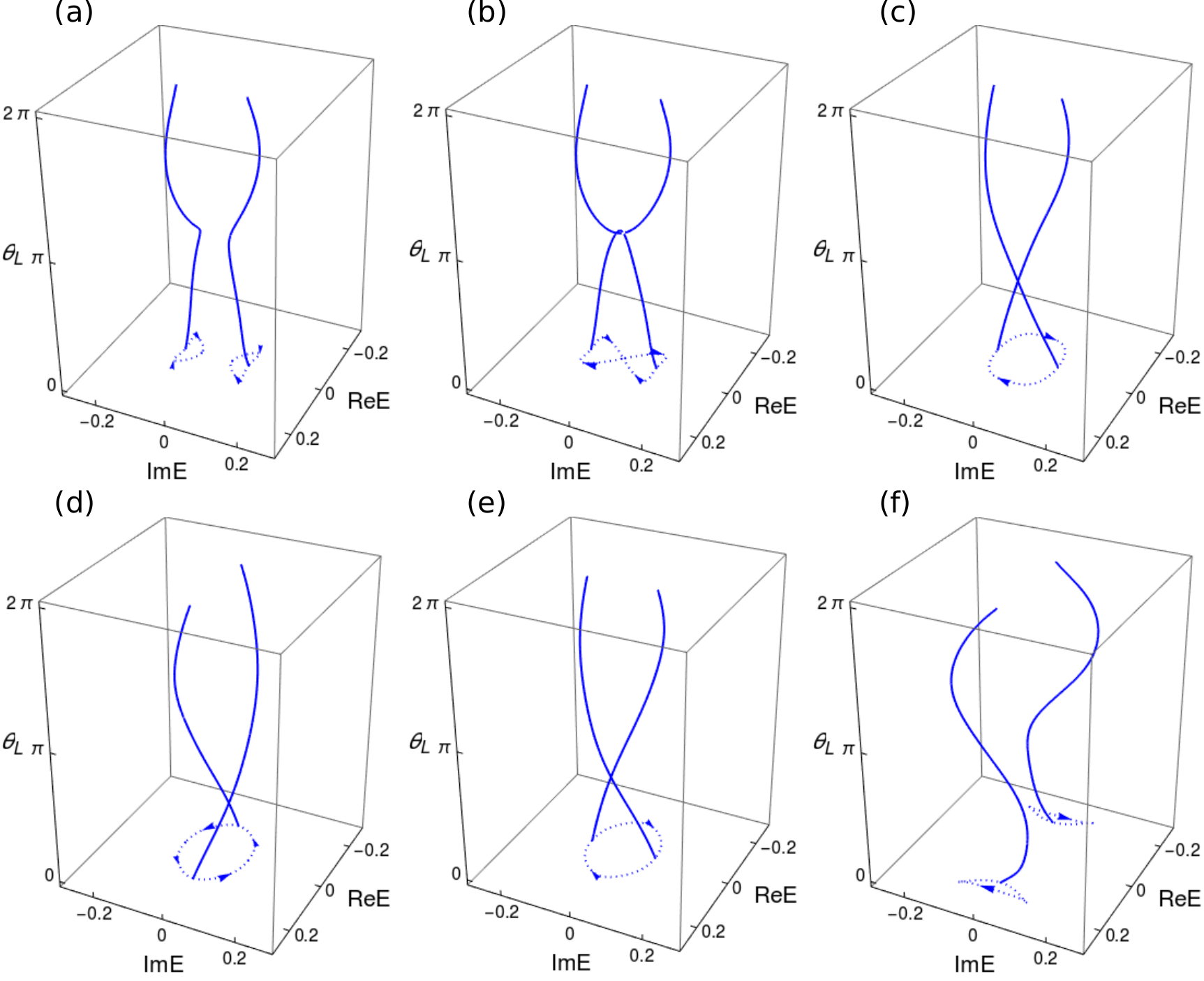}
  \caption{\textbf{Illustration of the topological phase transitions from vorticity.} The trajectory of the two complex eigenvalues, when the contour parameterized by $\theta_L \in [0,2\pi)$ (a) does not encircle any EP, (b) marginally touches the EP marked by $\gamma_1$, (c) encircles $\gamma_1$ completely, (d) encloses the EP marked by $\delta_1$, (e) encircles the EP marked by $\alpha_1$, and (f) encircles both the EPs marked as $\alpha_1$ and $\beta_1$. Note that in the case of the contour encircling both EPs $\alpha_1$ and $\beta_1$, the total vorticity vanishes as the two EPs have opposite vorticities. Their projections(dashed blue line) onto the complex plane are shown for a better view. The orientations of two eigenbands have been shown with the arrow. This clearly shows the swapping (or its lack thereof) of the complex energy values and the resulting vorticity. See Fig.~\ref{Location_of_EP} for labeling of the EPs.}  \label{Vorticity_2}
\end{figure*}

\subsection{Vorticity of energy eigenvalues}

 In addition to the winding number, there is another complementary topological invariant associated with the energy dispersion of non-Hermitian Hamiltonian, rather than the energy eigenstate. One can define, for any pair of the bands, the winding number of their energies $E_m(\textbf{k})$ and $E_n(\textbf{k})$ in the complex energy plane as follows~\cite{PhysRevLett.120.146402},
 
 \begin{equation}
     \nu_{mn}(\Gamma)=-\dfrac{1}{2\pi}{\oint_{\Gamma}{{\nabla_{\textbf k}} {\arg[E_m(\textbf k)-E_n(\textbf k)]}} \,d\textbf{k}},
 \end{equation}
 
where $\Gamma$ is a closed loop in momentum space. This is called the vorticity, $\nu_{mn}(\Gamma)$. We  write a non-Hermitian Hamiltonian of a periodic system in the parameter space of momentum $\textbf k$, whose eigenstates are Bloch waves and whose energies $E_n(\textbf{k})$ vary with momentum in the Brillouin zone. Here $m$ and $n$ are the band indices of different eigenstates. We define two complex energies, $E_m(\textbf{k}) \neq E_n(\textbf{k})$ for all $m \neq n$ and all \textbf{k}. For such a complex multi-band system, when the region of complex energies does not overlap with each other, i.e., $E_n(\textbf{k}) \neq E_m(\textbf{k'})$ for all $m \neq n$ and all $\textbf{k}$ and $\textbf{k'}$, then the band $E_n(\textbf{k})$ is found to be surrounded by a gap in the complex energy plane~\cite{PhysRevLett.120.146402,alvarez2018topological}. In this case we get zero vorticity. In another case, when we encircle an EP at $\textbf{k}_0$, at the band degeneracy point where $E_n(\textbf{k}_0) = E_m(\textbf{k}_0)$, we obtain two topologically different bands due to gap closing in the complex energy plane. In this case we find a non-zero vorticity. As a consequence of the square root singularity in the dispersion of Eq.~\ref{energy_eigenvalues}, both the pair of energy eigenvalues and the corresponding eigenstates are swapped as the momentum traverses along $\Gamma$ and we obtain the vorticity $\nu_{\Gamma}=1/2$~\cite{PhysRevLett.86.787,PhysRevE.69.056216,PhysRevLett.120.146402}. The Hamiltonian being non-Hermitian, we can in general write for a single complex band $E(k)=\mathopen|E(k)\mathclose| e^{i\theta(k)}$, where $\theta=\tan^{-1}({\mathrm{Im}E/\mathrm{Re}E})$~\cite{ghatak2019new}. We vary $\theta(k)$ in a periodic cycle $\theta(k)\rightarrow\theta(k)+2v\pi$ (where $v$ is an integer) without violating the periodicity. Whenever we enclose an EP along the real axis, we obtain a quantized vorticity. 
 
 For our semi-Dirac model, we discover three distinct cases by choosing different contours and their centers along the real axis by redefining $\theta(k)$ with respect to the base energy: 1) when we do not enclose an EP [Fig.~\ref{Vorticity_2}(a)], 2) when the contour marginally touches the EP [Fig.~\ref{Vorticity_2}(b)], and 3) when we enclose an EP [Fig.~\ref{Vorticity_2}(c)]. So, we clearly demonstrate a topological phase transition as the vorticity changes and there is a charateristic swapping of eigenvalues and corresponding eigenvectors as the momentum is traversed along the contour around an EP. As shown in Fig.~\ref{Vorticity_2}(c) and (d), for the two neighboring EPs the complex eigenvalues wind around each other in opposite directions, namely in clockwise and anticlockwise directions with vorticities $+1/2$ and $-1/2$, respectively. When we enclose odd number of EPs within the contour $\Gamma$, we obtain a half-integer vorticity, while for even number of EPs, the vorticity becomes an integer, as shown in Fig.~\ref{Vorticity_2}(f). These results are in agreement with our analysis in the previous subsection using the winding number.

\subsection{Bulk-boundary correspondence}

\begin{figure*}
\includegraphics[scale=0.35]{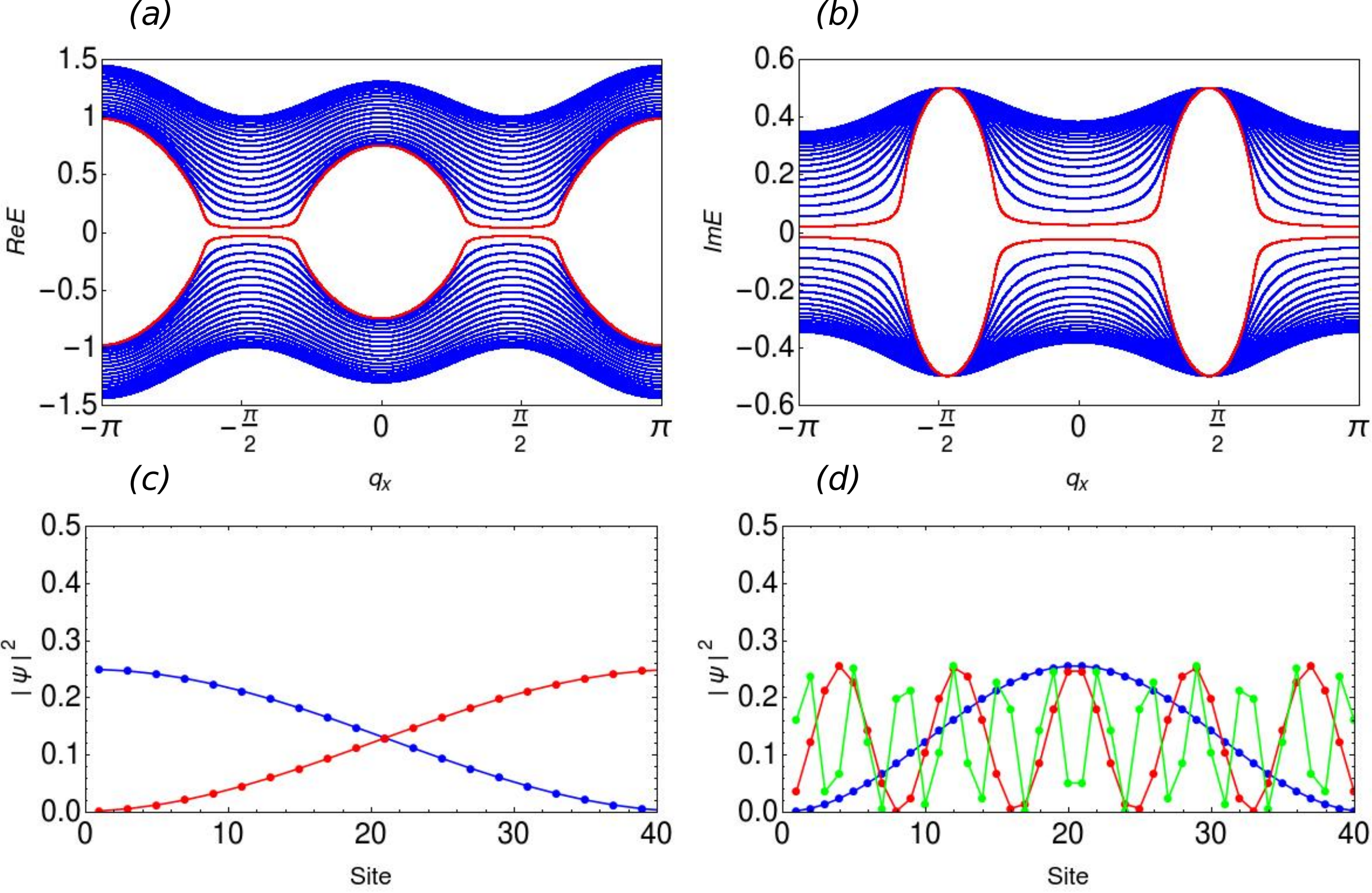}
 \caption{\textbf{Analysis of the lattice model under open boundary conditions presented in Eq.~\ref{lattice_hamiltonian_1}.} (a) The real part and (b) the imaginary part of energy eigenvalues are shown as a function of $q_x$. Note the presence of zero modes in both the real and imaginary part of the spectra in different ranges of $q_x$. However, they do not simultaneously go to zero, rendering the absolute value of the eigenvalue non-zero for all values of $q_x$. We use red colour to identify the eigenvalue corresponding to edge states. (c) The wave functions corresponding to the lowest energy states localized at opposite edges are shown as a function of position. Here we have chosen $q_x=0$. (d) The various wave functions corresponding to  bulk modes plotted along the lattice site index. We choose the following parameter values: $\gamma=0.5$, $v_f=1.0$, $m^\ast=1.0$ and $\delta_0=0.9$ with $N=40$ sites for the lattice model.}  \label{bulk-boundary_correspondence1}
\end{figure*}

An important aspect of topological systems is the bulk-boundary correspondence, where a topologically non-trivial bulk manifests in the form of protected boundary modes~\cite{RevModPhys.82.3045}. This feature and its modifications in non-Hermitian systems have received intense interest~\cite{alvarez2018topological,ghatak2019new,torres2019perspective,alvarez2018non,bergholtz2019exceptional}.Before we start analyzing lattice model we would like to clarify that an EP being the geometric property of the Hilbert space, its presence can be predicted in the perspective of the fidelity susceptibility~\cite{tzeng2020hunting}. Since there are only a limited number of $q = (q_x, q_y)$ points in a finite size system, the exceptional point may be absent in such a finite lattice model throughout the parameter space. It can be shown that the ground state fidelity susceptibility density characterizes an EP if one of the $q = (q_x, q_y)$ points is close to the $q_{EP}$~\cite{tzeng2020hunting}. To explore the bulk-boundary correspondence in our proposed semi-Dirac systems, we consider the following lattice Hamiltonian 

\begin{equation} \label{lattice_hamiltonian_1}
H=[1-m^\ast\cos({q_x})-\delta_0]\sigma_x+(v_f\sin{q_y}+i\gamma)\sigma_z.
\end{equation}

We note that this model exactly matches with the continuum model described in Eq.~\ref{Hamiltonian}. 

We make the replacements $q_i\rightarrow \sin{q_i}$ and $q_i^2\rightarrow 2(1-\cos{q_i})$ to obtain the above Hamiltonian starting from the continuum low-energy model. In the absence of non-Hermitian term $i\gamma\sigma_z$, the band degeneracy occurs along $q_y=0$ line, with the energy going to zero at $m_x=0$, where we define $m_x=1-m^\ast\cos({q_x})-\delta_0$. In the presence of the $i\gamma\sigma_z$ term, the condition for the bulk energy bands to touch the zero energy line is obtained to be $m_x=\pm\gamma$, similar to continuum model. On the practical side, this Hamiltonian can be obtained from the honeycomb lattice by tuning the hopping parameters appropriately~\cite{Montambaux2009,saha2016photoinduced}. In Appendix A, we present the analysis of several other models of semi-Dirac nature with different types of non-Hermitian terms.
 
To get more physical insights into the topological phase transitions accompanying the disappearance of zero energy modes, we begin with the tight-binding model consisting of two orbitals in the unit cell. For this model: $t_2=v_f/2$ represents the inter-cell and inter-orbital hopping, $m_x$ denotes the intra-cell and inter-orbital hopping, $i\gamma$ and $-i\gamma$ are the onsite gain and loss terms for the two orbitals (which we label by $A$ and $B$). For such a tight-binding model, the real space wave functions, $\psi_{A_{n}}$ and $\psi_{B_{n}}$, should satisfy

\begin{equation} \label{wavefunction3}
\begin{split}
i\gamma\psi_{A_n} + m_x\psi_{B_n} + it_2\psi_{A_{n-1}} - it_2\psi_{A_{n+1}} = E\psi_{A_n},\\
-i\gamma\psi_{B_n} + m_x\psi_{A_n} - it_2\psi_{B_{n-1}} + it_2\psi_{B_{n+1}} = E\psi_{B_n}.
\end{split}
\end{equation}
We choose the ansatz solution as $(\psi_{A_n},\psi_{B_n}) =\beta^n(\psi_A,\psi_B)$. Substituting these in Eq.~\ref{wavefunction3}, we obtain the coupled equations

\begin{equation} \label{wavefunction4}
\begin{split}
i[t_2(1/{\beta}-\beta)+\gamma]\psi_A   +m_x\psi_B= E\psi_A,\\
-i[t_2(1/\beta-\beta)+\gamma]\psi_B  +m_x\psi_A = E\psi_B.
\end{split}
\end{equation}

For non-trivial solutions, the above two equations yield the following condition

\begin{equation} \label{beta1-equation}
 t_2^2\beta^4 -2\gamma t_2 \beta^3+ (\gamma^2+E^2-m_x^2-2t_2^2)\beta^2 +2\gamma t_2 \beta+ t_2^2 = 0,
\end{equation}

which leads to four roots, $\beta_{i}$ ($i=1,2,3,4$). These roots satisfy $\beta_1 + \beta_2 + \beta_3 + \beta_4 = \dfrac{2\gamma}{t_2}$ and $\beta_1\beta_2\beta_3\beta_4=1$.\\ The four roots can be explicitly calculated to be 

\begin{equation}
\begin{split}
\beta_{1,2}=\dfrac{\gamma\pm\eta }{2t_2}-\dfrac{1}{2}\sqrt{2-\dfrac{\xi}{t_2^2}+\dfrac{2\gamma^2}{t_2^2}\pm \dfrac{2\gamma}{t_2^2}\eta}, \\
\beta_{3,4}=\dfrac{\gamma\mp \eta}{2t_2}+\dfrac{1}{2}\sqrt{2-\dfrac{\xi}{t_2^2}+\dfrac{2\gamma^2}{t_2^2}\mp \dfrac{2\gamma}{t_2^2}\eta},
\end{split}
\end{equation}

where we define $\eta=\sqrt{-\xi-2t_2^2+\gamma^2}$ and $\xi=\gamma^2+E^2-m_x^2-2t_2^2$. Next, let us consider a long chain. For the bulk eigenstates, two roots out of the four roots are required to satisfy the condition~\cite{PhysRevLett.121.086803,yang2019auxiliary}: $|{\beta_{2}}|=|{\beta_{3}}|=\beta$, leading to the constraint $\xi=0$. Considering the limit $E\rightarrow \pm \sqrt{2}t_2$, we get physically feasible solutions as $-\gamma \leq m_x \leq \gamma$ corresponding to the edge states. At these points, we find that $m_x(q_x)=\pm\gamma$. In this situation, the open boundary bulk spectra tend to touch zero energy, and there are accompanying topological phase transitions.

Next, we turn to numerical computations in order to complement our analytical results. We express our Hamiltonian in the parameter space of $q_x$ and choose open boundary conditions in the $y$ direction, along which the dispersion is linear. We choose $N=40$ sites along the $y$ direction and numerically compute the eigenvalues and eigenvectors. Here we consider right eigenvectors. For a non-Hermitian Hamiltonian one needs a bi-orthogonal coordinate system, such that $\sum_{j}\psi^{*}{_L}(x_j ) \psi_{R}(x_j )=1$, where $\psi^{*}{_L}(x_j )$ and $\psi_{R}(x_j )$ are the left and right eigenfunctions, respectively.  Our results are presented in Fig.~\ref{bulk-boundary_correspondence1} (also see Fig.~\ref{bulk-boundary_correspondence} in Appendix A). In general, we can divide the eigenspectra into two qualitatively different regions:

1) The region where $\mathrm{Re}[E]\rightarrow 0$ and $\mathrm{Im}[E]\neq 0$. In such a case, we obtain states localized at the edges, corresponding to energy eigenvalues satisfying $E=i\gamma$. All the other eigenstates are bulk-like in nature. This is the $PT$ symmetry broken region between $-\gamma \leq m_x \leq \gamma$, which is consistent with our analytical calculation.

2)The region corresponding to $\mathrm{Im}[E]\rightarrow 0$ and $\mathrm{Re}[E]\neq 0$. In this contrasting case, we obtain localized edge states corresponding only to the lowest energy eigenvalues ($E\rightarrow \pm \sqrt{2}t_2$) and bulk states corresponding to the rest of the energy eigenvalues. This is the $PT$ symmetry unbroken case.

We find edge modes in the region of $\mathrm{Im}[E]=0$ and $\mathrm{Re}[E]\neq 0$, between the two Dirac points. We get a superposition of two edge modes localized at the left and right edges, corresponding to lowest energy eigenvalue. The left edge mode (blue) occupies the odd states while the right edge mode (red) occupies the even sites. As previously mentioned, these correspond to the lowest energy eigenstate. Upon introducing the non-Hermitian term, we create two EPs originating out of each semi-Dirac point in the region between $-\gamma \leq m_x \leq \gamma$. In between each pairs of EPs, we do not find any edge states corresponding to the lowest energy eigenvalue. Indeed from our continuum model analysis, we obtain exactly the same features. The winding number turns out to be zero between each pairs of EPs (see the phase diagram in Fig.~\ref{Topological_phase_Transition}). Outside this region the winding number is non-zero and we do find edge modes. So, we conclude that our non-Hermitian semi-Dirac model shows a bulk-boundary correspondence, in marked contrast to several other non-Hermitian topological models.

An associated anomalous feature is the non-Hermitian skin effect, where a macroscopic number of eigenvectors become localized at the boundaries~\cite{PhysRevLett.121.086803}. Since, our model shows bulk-boundary correspondence, we expect the non-Hermitian skin effect to be absent in semi-Dirac semimetals. Indeed, for our model, we find that there are no general values of $\beta$, such that $\beta<1$ or $\beta>1$, and we do not obtain a macroscopic number of eigenvectors that are localized at the left or right boundaries.

\subsection{Effect of asymmetric hopping on non-Hermitian skin effect}
In the non-Hermitian setting, the spectral properties are sensitive to boundary condition with a topological distinction between the model with PBC and OBC leading to a violation of the bulk-boundary correspondence~\cite{bergholtz2019exceptional}. Under OBC the associated eigenfunctions take the following general form

\begin{equation}
    \psi(x)=\beta^x u(x).
\end{equation}\\

Unless $|\beta|=1$, such a wavefunction tends to be localized at the edge of the system resulting in the non-Hermitian skin effect. Thus non-Hermitian skin effect is typically found in system with non-reciprocal hopping~\cite{alvarez2018topological,kunst2018biorthogonal}. To study the effect of asymmetric hopping on non-Hermitian topological phases in terms of the non-Bloch band theory, we first briefly discuss the Hatano-Nelson model under open boundary condition~\cite{imura2020generalized,yokomizo2019non}, before analysing the semi-Dirac case. The Hamiltonian for the Hatano-Nelson model reads

\begin{equation}
    H=\sum_{i=1}^{L-1}\left((t+g)c^{\dagger}_{i+1 } c_{i} +(t-g)c^{\dagger}_{i } c_{i+1}\right).
\end{equation}

We assume $t>|g| \geq 0$ for the sake of simplicity, where $g$ introduces an asymmetry in left and right hopping and $L$ is the system size. Consider the eigenvector $|\psi\rangle = (\psi_1,\psi_2,....\psi_L)^T$, which satisfies the real space eigenvalue equation

\begin{equation}
    H|\psi\rangle=E|\psi\rangle.
\end{equation}

The wavefunction $\psi$ is subjected to the following boundary condition

\begin{equation}\label{bc}
    \psi_0=\psi_{L+1}=0.
\end{equation}

Now, we consider a generic eigenstate as an ansatz solution given as

\begin{equation}
    \psi_i=c_{+} \beta_{+}^{i}+c_{-} \beta_{-}^{i},
\end{equation}

where $\beta_{j}$ $(j= +,-)$ are the solutions of the eigenvalue equation with coefficients $c_{+},c_{-}$ which are in general complex. The corresponding eigenenergy $E(\beta)$ is determined as

\begin{equation}\label{eg}
    (t+g) \beta_{j}^{-1} + (t-g) \beta_j =E.
\end{equation}

Now using the boundary condition in Eq.~\ref{bc} we have

\begin{equation}\label{bc1}
\psi_0=0 \implies c_{+} + c_{-} =0,
\end{equation}

\begin{equation}\label{bc2}
\psi_{L+1}=0 \implies \left(\dfrac{\beta_{+}}{\beta_{-}}\right)^{L+1}=1.
\end{equation}

In general Eq.~\ref{bc2} can be written as

\begin{equation}
    \dfrac{\beta_{+}}{\beta_{-}}=e^{2i\phi_m},
\end{equation}

where $\phi_m=\dfrac{m\pi}{L+1}$ $(m=1,..L)$. Ultimately we obtain $|\beta_{+}|=|\beta_{-}|$. Following Eq.~\ref{eg} we get the condition

\begin{equation}
   r= |\beta_{+}|=|\beta_{-}|=\sqrt{\dfrac{t+g}{t-g}},
\end{equation}

where $r$ represents the degree of skin effect under OBC. Eq. \ref{bc1} indicates that the wavefunction takes the form

\begin{equation}
     \psi_{i} \propto (\beta_{+}^{i}-\beta_{-}^{i}).
\end{equation}

As $r\neq 1$ the wavefunction tends to damp or amplify exponentially towards the end of the system resulting in the non-Hermitian skin effect~\cite{imura2020generalized,yokomizo2019non,yao2018edge}.
 
Having illustrated the effect of asymmetric hopping using the Hatano-Nelson model, we next move to the discussion of our semi-Dirac case. All the models having semi-Dirac dispersion discussed in the main text and the appendix do not show the non-Hermitian skin effect. However, one can think of other models with semi-Dirac dispersion that can show this non-Hermitian skin effect. Crucially, the non-Hermitian skin effect not only depends on dispersion but also on asymmetric hopping~\cite{yokomizo2019non}, as we have seen. The following model~\cite{yao2018edge} with $t_2$ nonzero in a specific parameter region shows the non-Hermitian skin effect upon choosing open boundary conditions in the $q_y$ direction. However, if we consider the model with $t_2=0$, to get back our original model, and the non-Hermitian skin effect is absent. Following ~\cite{yokomizo2019non,yao2018edge}, the Hamiltonian can be written as 

\begin{equation}
    H(q)=(m_x-t_2\cos{q_y})\sigma_x+(v\sin{q_y+i\gamma})\sigma_y,
\end{equation}

where $m_x=M-t_1\cos{q_x}$ and we further choose $t_2=v$ to obtain a quadratic equation for $\beta$ following the procedure outlined earlier.

\begin{equation}
    \beta^2(m_x+\gamma)v+\beta(E^2+\gamma^2-v^2-m_x^2)+(m_x-\gamma)v=0,
\end{equation}

leading to two solutions, $\beta_1$ and $\beta_2$ , which satisfy

\begin{equation}
    \beta_1 \beta_2=\dfrac{m_x-\gamma}{m_x+\gamma}
\end{equation}

For a long chain, $|\beta_1|=|\beta_2|$ leads to 

\begin{equation}
    |\beta_j|=\sqrt{\dfrac{m_x-\gamma}{m_x+\gamma}}.
\end{equation}
 
 When $|\beta_j|<1$ $(|\beta_j|>1)$ we obtain all the bulk states left (right) localized leading to the non-Hermitian skin effect. However, in contrast, for our case with $t_2=0$, we do not find any generalized $\beta$ such that $\beta<1$ or $\beta>1$ in the parameter space. Further, the product of four roots turns out to be unity, and hence we find no non-Hermitian skin effect.

\section{Summary and conclusions}

In closing, we would like to briefly discuss possible experimental realization of our proposed non-Hermitian semi-Dirac semimetals. Several ingenious experimental setups have been suggested as well as realized, to engineer non-Hermitian topological semimetals. These include cold atoms~\cite{xu2017weyl}, waveguides~\cite{cerjan2019experimental} and microwave cavity experiments~\cite{chen2017exceptional}, to highlight just a few. Many recent experiments have been devoted to study the topological behaviour of exceptional points and their topological phase transitions~\cite{dembowski2001experimental,doppler2016dynamically,ding2016emergence}. Very recently, Weyl exceptional rings have been proposed in the cold atomic gas trapped in optical lattice system upon introducing particle gain and loss perturbations~\cite{xu2017weyl}. In an exciting recent work by Cerjan \textit{et al.}~\cite{cerjan2019experimental}, non-Hermitian Weyl exceptional rings were created in an evanescent-coupled bipartite optical waveguide array and their topological transitions were demonstrated in a controllable manner. Our proposal is amenable to realization in such cold atom and waveguide setups.

To summarize, we proposed a new class of non-Hermitian semi-Dirac semimetals, in the presence of particle gain and loss perturbations. We showed that a non-Hermitian term creates exceptional points in the spectrum, emerging out of each of the semi-Dirac points, which have distinct topological signatures. We illustrate the topological phase transitions by evaluating two different topological markers and map out the complete phase diagram for our model. We examined the nature of the edge states and bulk-boundary correspondence by using numerical and analytical calculations. Interestingly, we discover that the non-Hermitian semi-Dirac semimetal admits both the presence and absence of non-Hermitian skin effect for different classes of models. We are hopeful that our findings will motivate further theoretical and experimental studies of these intriguing topological systems.

\section*{Acknowledgments}

A.B. would like to acknowledge the Indian Institute of Science for a fellowship. A.N. acknowledges support from the start-up grant (SG/MHRD-19-0001) of the Indian Institute of Science and and DST-SERB (project number SRG/2020/000153).

\section{APPENDIX A: Analysis of different non-Hermitian models}

\begin{figure*}
\includegraphics[scale=0.35]{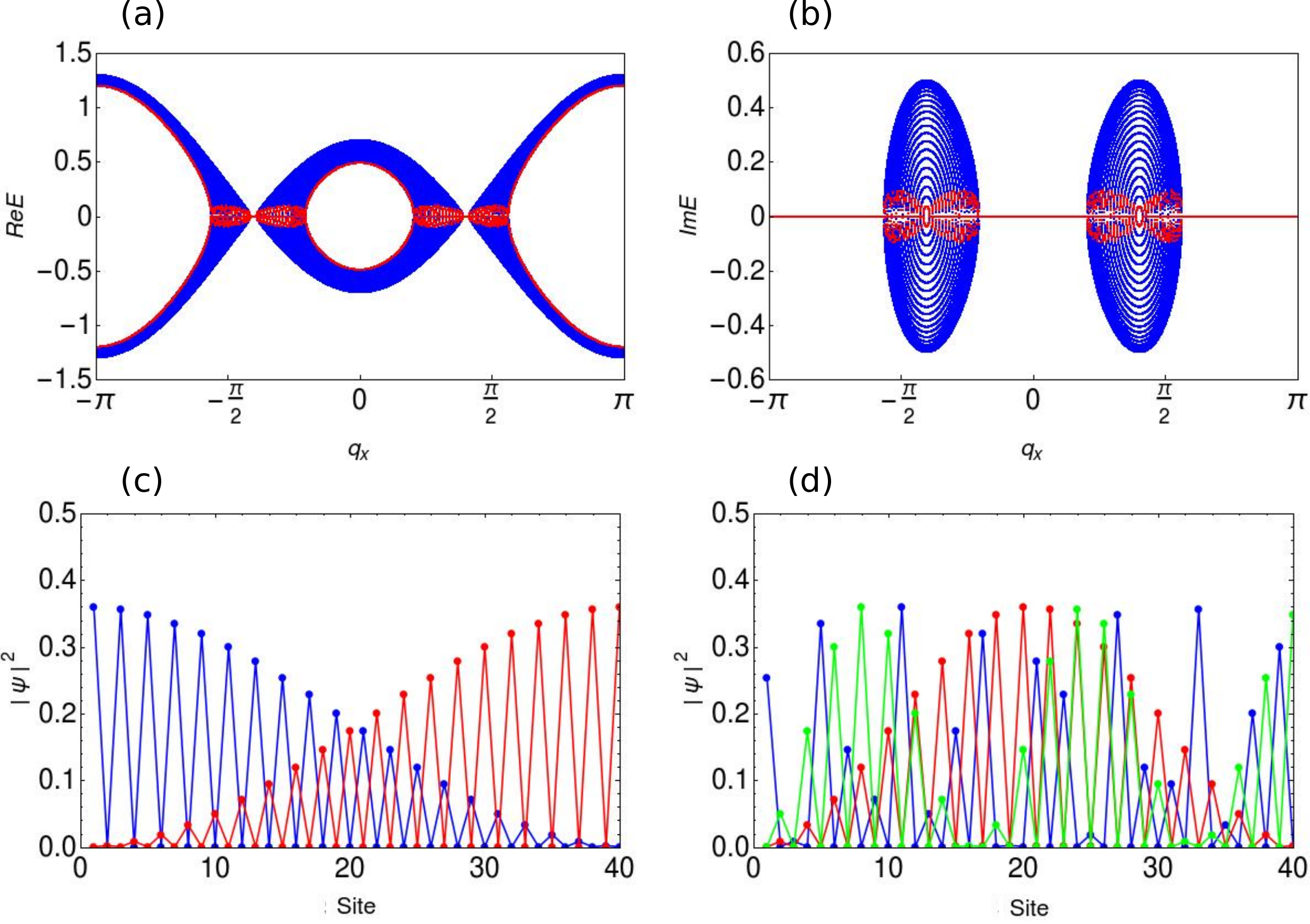}
  \caption{\textbf{Analysis of the lattice model under open boundary conditions corresponding to Eq.~\ref{lattice_hamiltonian} } (a) The real part and (b) the imaginary part of energy eigenvalues are shown as a function of $q_x$. Note the presence of zero modes in both the real and imaginary part of the spectra in different ranges of $q_x$. However, they do not simultaneously go to zero, rendering the absolute value of the eigenvalue non-zero for all values of $q_x$. We use red colour to identify the eigenvalue corresponding to edge states. (c) The wave functions corresponding to the lowest energy states localized at opposite edges are shown as a function of position. Here we have chosen $q_x=0$. (d) The various wave functions corresponding to  bulk modes plotted along the lattice site index. Here we have chosen various eigenstates corresponding to the highest, 10th and 30th energy eigenvalue, at $q_{x}=0$. We choose the following parameter values: $\gamma=0.5$, $v_f=0.5$, $m^\ast=1.0$ and $\delta_0=0.7$ with $N=40$ sites for the lattice model.}  \label{bulk-boundary_correspondence}
\end{figure*}

\begin{figure*}
\includegraphics[scale=0.35]{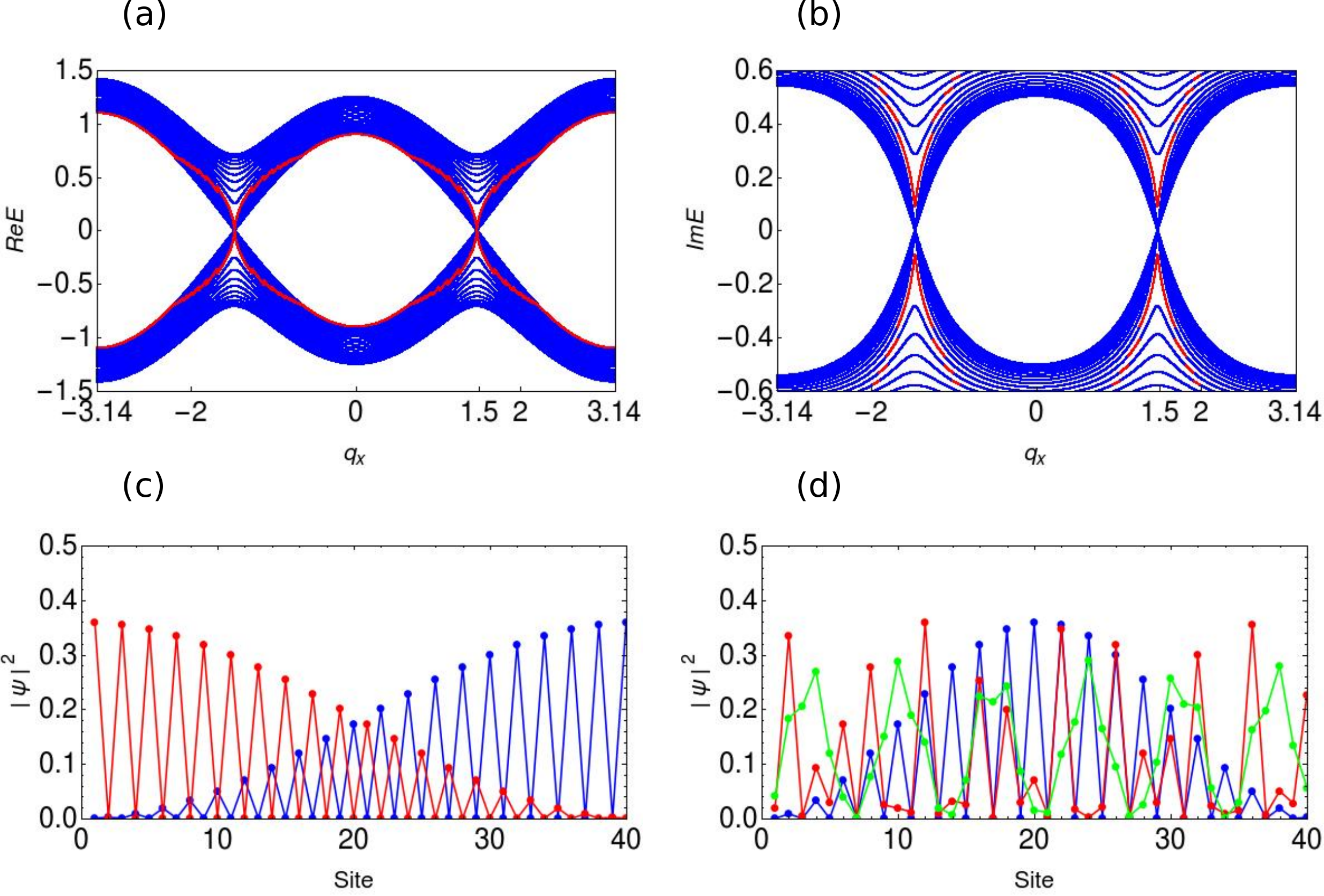}
 \caption{\textbf{Analysis of the lattice model corresponding to Eq.~\ref{lattice_hamiltonian_2} under open boundary conditions.} (a) The real part and (b) the imaginary part of energy eigenvalues are shown as a function of $q_x$. Note the presence of zero modes in both the real and imaginary part of the spectra in different ranges of $q_x$. However, they do not simultaneously go to zero, rendering the absolute value of the eigenvalue non-zero for all values of $q_x$. We use red colour to identify the eigenvalue corresponding to edge states. (c) The wave functions corresponding to the lowest energy states localized at opposite edges are shown as a function of position. Here we have chosen $q_x=0$. (d) The various wave functions corresponding to  bulk modes plotted along the lattice site index. We choose the following parameter values: $\gamma=0.5$, $v_f=0.5$, $m^\ast=1.0$ and $\delta_0=0.7$ with $N=40$ sites for the lattice model.}  \label{bulk-boundary_correspondence3}
\end{figure*}

In this Appendix, we will study different semi-Dirac models with various kinds of non-Hermitian terms. We will present numerical results on these under open boundary conditions, as well as analytical results on the nature of bulk-boundary correspondence.\\

We first consider the following tight binding model consisting of a basis of two orbitals in a unit cell

\begin{equation} \label{lattice_hamiltonian}
H=[1-m^\ast\cos({q_x})-\delta_0]\sigma_x+v_f\sin{q_y}\sigma_y+i\gamma\sigma_z.
\end{equation}

For this model: $it_1$ with $t_1=v_f/2$ represents the intercell and intraorbital hoppings for the $A$ and $B$ orbitals, $m_x$ denotes intracell and inter-orbital hopping, $i\gamma$ and $-i\gamma$ are the onsite gain and loss terms for the two orbitals. For this model, the real space wave functions, $\psi_{A_{n}}$ and $\psi_{B_{n}}$, should satisfy

\begin{equation} \label{wavefunction1}
\begin{split}
i\gamma\psi_{A_n} + m_x\psi_{B_n} - t_1\psi_{B_{n-1}} + t_1\psi_{B_{n+1}} = E\psi_{A_n},\\
-i\gamma\psi_{B_n} + m_x\psi_{A_n} + t_1\psi_{A_{n-1}} - t_1\psi_{A_{n+1}} = E\psi_{B_n}.
\end{split}
\end{equation}

We choose the ansatz solution as $(\psi_{A_n},\psi_{B_n}) =\beta^n(\psi_A,\psi_B)$. Substituting these in Eq.~\ref{wavefunction1}, we obtain the coupled equations

\begin{equation} \label{wavefunction2}
\begin{split}
[t_1(\beta-1/{\beta})+m_x]\psi_B + i\gamma\psi_A = E\psi_A,\\
[t_1(1/\beta-\beta)+m_x]\psi_A -i\gamma\psi_B = E\psi_B.
\end{split}
\end{equation}

For non-trivial solutions, the above two equations yield the following condition

\begin{equation} \label{beta-equation}
 t_1^2\beta^4 + (\gamma^2+E^2-m_x^2-2t_1^2)\beta^2 + t_1^2 = 0,
\end{equation}

which leads to four roots, $\beta_{i}$ ($i=1,2,3,4$). These four roots satisfy a very similar condition as we get from Eq.~\ref{beta1-equation} in the main text. These roots satisfy $\beta_1 + \beta_2 + \beta_3 + \beta_4 = 0$ and $\beta_1\beta_2\beta_3\beta_4=1$. The four roots can be explicitly calculated to be 

\begin{equation}
\begin{split}
\beta_{1,2}=\pm\sqrt{\frac{\zeta+\sqrt{\zeta^2-4}}{2}}, \\
\beta_{3,4}=\pm\sqrt{\frac{\zeta-\sqrt{\zeta^2-4}}{2}},
\end{split}
\end{equation}

where we define $\zeta=\frac{m_x^2+2t_1^2-\gamma^2-E^2}{t_1^2}$. Next, let us consider a long chain. For the bulk eigenstates, the four roots are required to satisfy the condition~\cite{PhysRevLett.121.086803,yang2019auxiliary}: $|{\beta_{2}}|=|{\beta_{3}}|=\beta$, leading to the constraint $\zeta=\pm 2$. Taking the $E\rightarrow 0$ limit of this $\zeta=\pm 2$ condition, we get physically feasible solutions as $-\gamma \leq m_x \leq \gamma$. At these points, we find that $m_x(q_x)=\pm\gamma$. In this situation, the open boundary bulk spectra touch zero energy, and there are accompanying topological phase transitions.

We have confirmed this by a numerical analysis of the model under open boundary conditions. We present the eigenspectra for $N=40$ sites ribbon in Fig.~\ref{bulk-boundary_correspondence}. We note that the features are similar in nature to that obtained for the model in Eq.~\ref{lattice_hamiltonian_1} of the main text. In particular, our examination of several randomly chosen bulk states reveals the absence of non-Hermitian skin effect. For this model also, we can divide the eigenspectra into two qualitatively different regions presented in the main text. Observing the nature of edge states we also find that these are in good agreement with bulk topological invarinat and this model obeys bulk-boundary correspondence.

Since this model shows bulk-boundary correspondence, we also expect the non-Hermitian skin effect to be absent for this model and we find that there do not exist generalized $\beta$ such that $\beta<1$ or $\beta>1$, and we do not obtain a macroscopic number of eigenvectors that are localized at the left or right boundaries. A complementary point of view to understand the absence of the non-Hermitian skin effect in our model is to use the saddle-point criterion, which was very recently introduced by Longhi~\cite{longhi2019probing}, based on the geometrical concept of generalized Brillouin zone ~\cite{yokomizo2019non,yao2018non,yang2019auxiliary}.
Longhi's criterion states that if there exists at least one saddle point of $Q(\Tilde{\beta})=E^2$ that does not lie on the unit circle $C_{\Tilde{\beta}}$, then a non-Hermitian Hamiltonian will show the non-Hermitian skin effect. Here $\Tilde{\beta}=e^{iq}$. For our Hamiltonian given in Eq.~\ref{lattice_hamiltonian} we construct the following equation in the parameter space of $q_x$

\begin{equation}
\begin{split}
Q(\Tilde{\beta})= m^2_x - v^2_f\left(\dfrac{\Tilde{\beta}^2-1}{2\Tilde{\beta}}\right)^2-\gamma^2.
\end{split}
\end{equation}

We obtain the saddle points of $Q(\Tilde{\beta})$ to be located at $\Tilde{\beta}=\pm1$ for any value of $m_x$ and $\gamma$. So, in general, the two saddle points lie on the unit circle spanned by $\Tilde{\beta}$. Therefore, according to Longhi's saddle point criterion, the model does not display a non-Hermitian skin effect, consistent with our numerical analysis.

We next present two more models with different forms of the non-Hermitian terms. \\

First, we consider the following model

\begin{equation} \label{lattice_hamiltonian_2}
H=[1-m^\ast\cos({q_x})-\delta_0+i\gamma]\sigma_x+v_f\sin{q_y}\sigma_z,
\end{equation}

with gain-loss term now proportional to $\sigma_x$. The tight binding model describing this Hamiltonian is similar to previous models, except for the form of the gain-loss term. The real space wavefunctions, $\psi_{A_{n}}$ and $\psi_{B_{n}}$, satisfy

\begin{equation} \label{wavefunction5}
\begin{split}
(m_x+i\gamma)\psi_{B_n} + it_2\psi_{A_{n-1}} - it_2\psi_{A_{n+1}} = E\psi_{A_n},\\
( m_x+i\gamma)\psi_{A_n} - it_2\psi_{B_{n-1}} + it_2\psi_{B_{n+1}} = E\psi_{B_n}.
\end{split}
\end{equation}

Using the ansatz solution as employed in the main text and substituting these in Eq.~\ref{wavefunction5}, we ultimately end up with the following condition on $\beta$

\begin{equation} \label{beta2-equation}
 t_2^2\beta^4 + (E^2-2t_2^2-m_x^2+\gamma^2-2im_x\gamma)\beta^2 + t_2^2 = 0,
\end{equation}

which leads to four roots, $\beta_{i}$ ($i=1,2,3,4$). These four roots satisfy the same condition as we get from Eq.~\ref{beta-equation}  in the main text: $\beta_1 + \beta_2 + \beta_3 + \beta_4 = 0$ and $\beta_1\beta_2\beta_3\beta_ 4=1$.
For a sufficiently long chain, $|{\beta_{2}}|=|{\beta_{3}}|=\beta$, and the model does not exhibit a non-Hermitian skin effect as the product of four roots is unity. 
Furthermore we analyzed the properties of this model numerically as presented in Fig.~\ref{bulk-boundary_correspondence3}. Consistent with our analytical results, the model shows an absence of non-Hermitian skin effect. 

\par If we choose off-diagonal terms to be different, leading to a different form of non-Hermiticity, the  Hamiltonian in Eq.~\ref{lattice_hamiltonian_2} becomes

\begin{equation} \label{lattice_hamiltonian_4}
H=[1-m^\ast\cos({q_x})-\delta_0]\sigma_x+v_f\sin{q_y}\sigma_z+
\begin{pmatrix} 
0 & \gamma_1 \\
\gamma_2 & 0 
\end{pmatrix},
\end{equation}

The Eq.~\ref{wavefunction5} then reads 

\begin{equation} \label{wavefunction6}
\begin{split}
(m_x+i\gamma_1)\psi_{B_n} + it_2\psi_{A_{n-1}} - it_2\psi_{A_{n+1}} = E\psi_{A_n},\\
( m_x+i\gamma_2)\psi_{A_n} - it_2\psi_{B_{n-1}} + it_2\psi_{B_{n+1}} = E\psi_{B_n}.
\end{split}
\end{equation}

The condition on $\beta$ from Eq.~\ref{beta2-equation} becomes
\begin{equation} \label{beta3-equation}
 t_2^2\beta^4 + (E^2-2t_2^2-m_x^2+\gamma_1\gamma_2-im_x(\gamma_1+\gamma_2))\beta^2 + t_2^2 = 0.
\end{equation}

This equation leads to the same conditions  $\beta_1 + \beta_2 + \beta_3 + \beta_4 = 0$ and $\beta_1\beta_2\beta_3\beta_ 4=1$. So, this analysis shows that this model too does not exhibit the non-Hermitian skin effect.\\



%

\end{document}